\documentclass[final]{svjour3}
\usepackage[dvipdfmx]{graphicx}
\usepackage{subcaption}
\usepackage{rotating}
\usepackage{amssymb}
\usepackage{mathptmx}
\usepackage{xspace}
\usepackage{marvosym}
\usepackage[numbers]{natbib}
\usepackage{xcolor}
\makeatletter
\journalname{Journal of Low Temperature Physics}
\newcommand{\umux}{$\mu$mux\xspace}

\newcommand{\affl}[1]{$^{#1}$}

\bibpunct{}{}{,}{s}{}{,}

\begin{document}

\newcommand{\hdblarrow}{H\makebox[0.9ex][l]{$\downdownarrows$}-}
\title{Bandwidth and Aliasing in the Microwave SQUID Multiplexer}

\author{C. Yu\affl{1} \and Z. Ahmed\affl{2} \and J.A. Connors\affl{3,4} \and J.M. D'Ewart\affl{2} \and B. Dober\affl{3} \and J.C. Frisch\affl{2} \and S.W. Henderson\affl{2} \and G.C. Hilton\affl{3} \and J. Hubmayr\affl{3} \and S.E. Kuenstner\affl{1} \and J.A.B. Mates\affl{3} \and M. Silva-Feaver\affl{5} \and J.N. Ullom\affl{3} \and L.R. Vale\affl{3} \and D. Van Winkle\affl{2} \and E. Young\affl{1}}

\institute{1. Department of Physics, Stanford University, Stanford, CA 94303\\
2. SLAC National Accelerator Laboratory, Menlo Park, CA 94025\\
3. National Institute of Standards and Technology, Boulder CO 80309\\
4. Department of Physics, University of Colorado Boulder, Boulder CO 80303\\
5. Department of Physics, University of California San Diego, La Jolla CA 92092\\
\Letter\, \texttt{cyndiayu@stanford.edu}}

\maketitle

\begin{abstract}

The microwave SQUID multiplexer (\umux) has enabled higher bandwidth or higher channel counts across a wide range of experiments in particle physics, astronomy, and spectroscopy. 
The large multiplexing factor coupled with recent commercial availability of microwave components and warm electronics readout systems make it an attractive candidate for systems requiring large cryogenic detector counts. 
Since the multiplexer is considered for both bolometric and calorimetric applications across several orders of magnitude of signal frequencies, understanding the bandwidth of the device and its interaction with readout electronics is key to appropriately designing and engineering systems. 
Here we discuss several important factors contributing to the bandwidth properties of \umux systems, including the intrinsic device bandwidth, interactions with warm electronics readout systems, and aliasing. 
We present simulations and measurements of \umux devices coupled with SLAC Microresonator RF (SMuRF) tone-tracking electronics and discuss several implications for future experimental design. 
\keywords{multiplexing, microwave SQUIDs, bandwidth}

\end{abstract}

\section{Introduction}

As cryogenic detector arrays become easier to fabricate and more desirable for low noise experiments, cryogenic multiplexing is increasingly important. 
Microwave resonator-based multiplexing techniques are increasingly popular due to high multiplexing factor potential, commercial availability of cryogenic components, and new advances in warm readout electronics. 
In particular, the microwave SQUID has emerged as a popular cold multiplexer for a variety of applications, including but not limited to cosmic microwave background polarimetry\cite{ari20, so18}, x-ray astronomy\cite{lynx_concept}, spectroscopy\cite{mates17,bennett19}, and neutrino calorimetry\cite{holmes18}.
The wide range of applications across many orders of magnitude of signal frequency necessitate careful consideration of the bandwidth properties of the multiplexer device in order to appropriately match the detectors to the readout. 
The system bandwidth is additionally impacted by the choice of readout electronics; in particular we consider here the impact of tone-tracking\cite{smurf18}, which allow for improvements in linearity and signal to noise.

\section{The Microwave SQUID Multiplexer}

\begin{figure}[htbp]
\begin{center}
\includegraphics[width=0.8\linewidth, keepaspectratio]{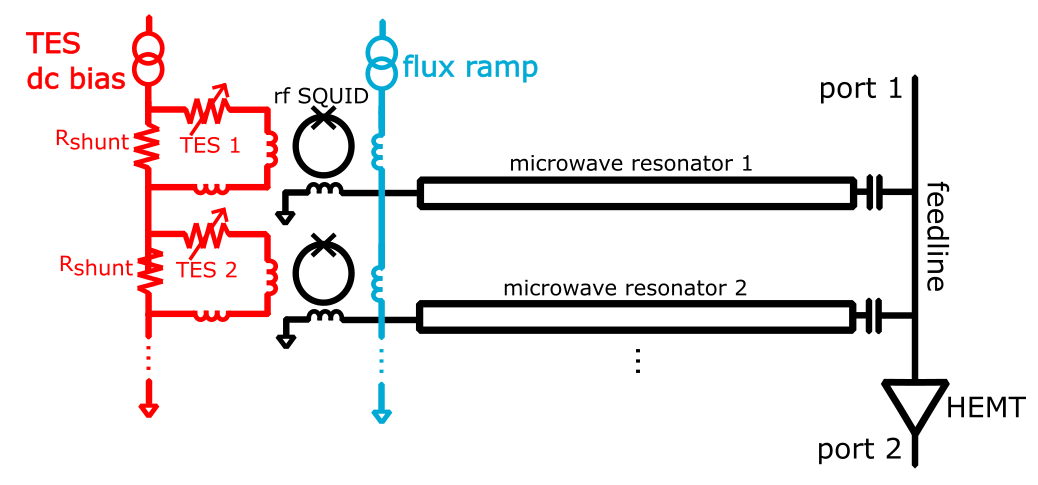}
\caption{(Color online) A schematic of the microwave SQUID multiplexer, shown here with two transition-edge sensors as the detector elements. Each detector is inductively coupled to an rf SQUID, which is in turn coupled to a microwave resonator on a common feedline. Subject to bandwidth constraints, up to several thousand resonators may be placed on the same feedline. A flux ramp linearizes the response.}
\label{fig:umuxschematic}
\end{center}
\end{figure}

The \umux device consists of a GHz-frequency resonator inductively coupled to a unique rf SQUID such that modulation of SQUID flux results in modulation of the resonance frequency.\cite{irwin04,mates08} 
A detector, most commonly a transition-edge sensor or magnetic microcalorimeter\cite{wegner18,dober21} is inductively coupled to a single corresponding SQUID. 
By tuning the resonators to have unique resonance frequencies, the detectors may be multiplexed by monitoring many resonances simultaneously for frequency shifts. 
A common flux ramp line, typically carrying a sawtooth pattern, linearizes the SQUID response such that detector signal appears as changes in the phase of the flux ramp modulated signal.\cite{mates08}
The system combines the heritage of dc-biased detectors with the multiplexing potential of GHz-resonators commonly found in KID systems.\cite{s4tech}
A schematic is given in Fig.~\ref{fig:umuxschematic}.

\begin{figure}[htbp]
     \centering
     \begin{subfigure}[b]{0.32\textwidth}
         \centering
         \includegraphics[width=\textwidth]{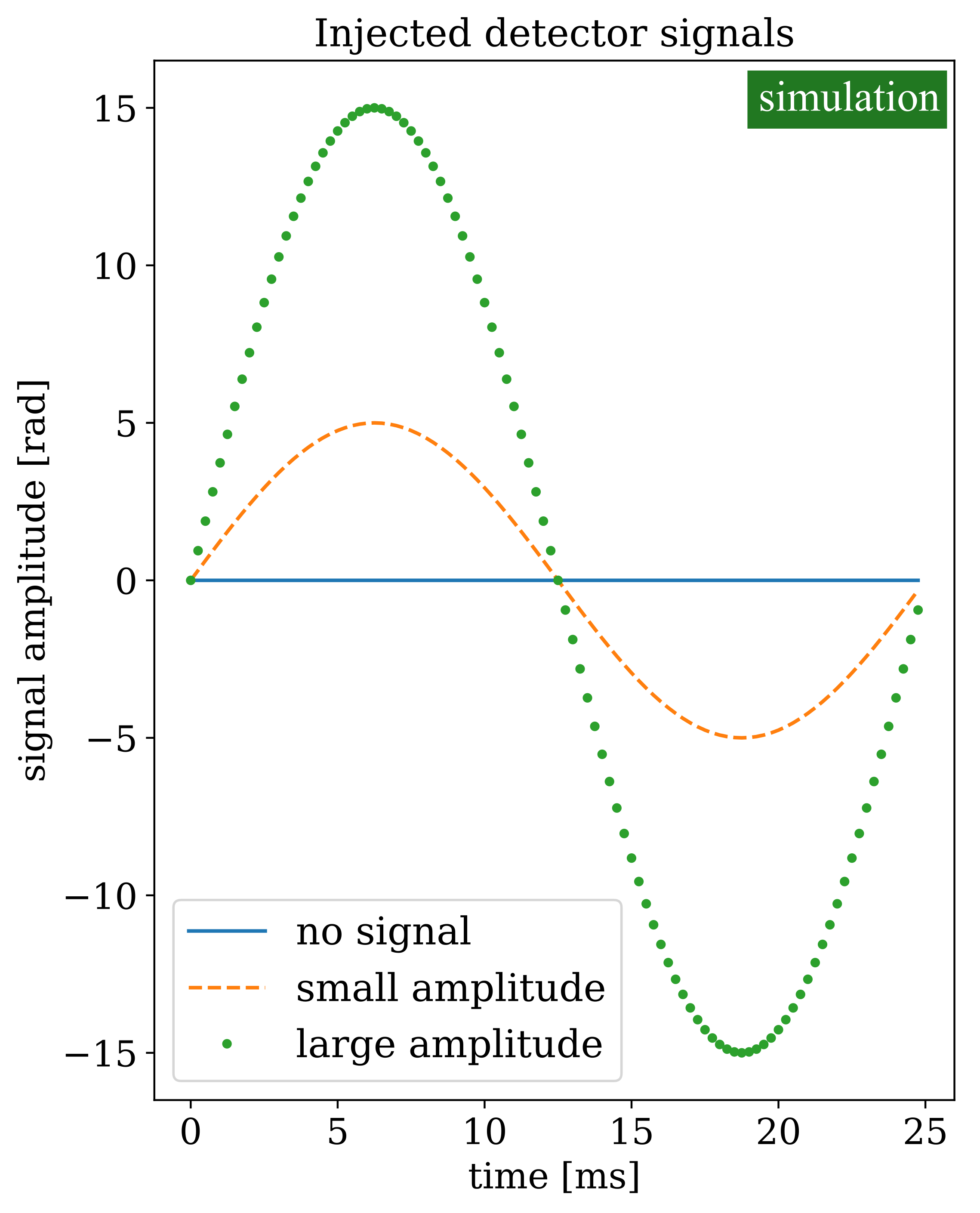}
         \label{fig:detsig}
     \end{subfigure}
     \hfill
     \begin{subfigure}[b]{0.31\textwidth}
         \centering
         \includegraphics[width=\textwidth]{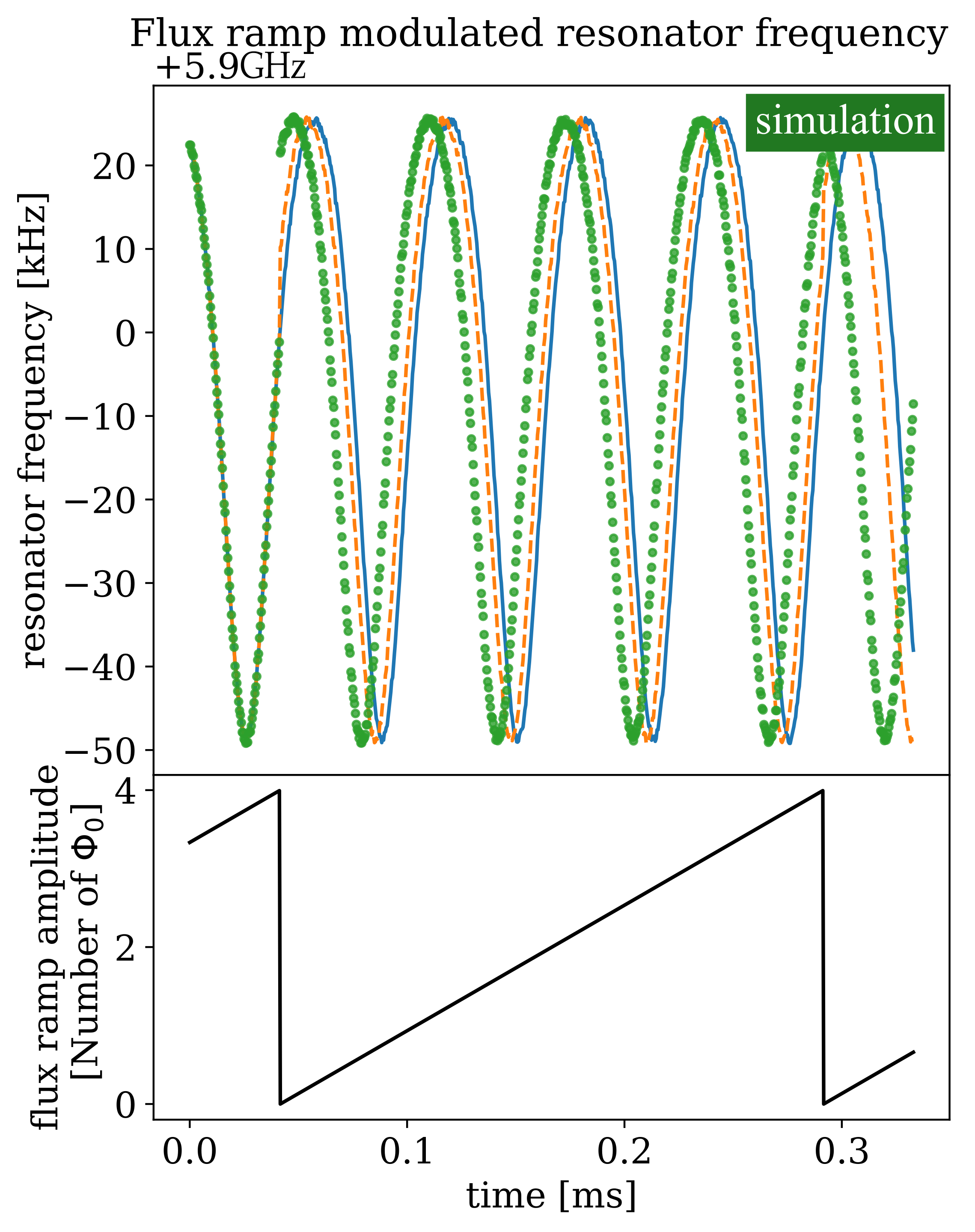}
         \label{fig:resfreq_mod}
     \end{subfigure}
     \hfill
     \begin{subfigure}[b]{0.32\textwidth}
         \centering
         \includegraphics[width=\textwidth]{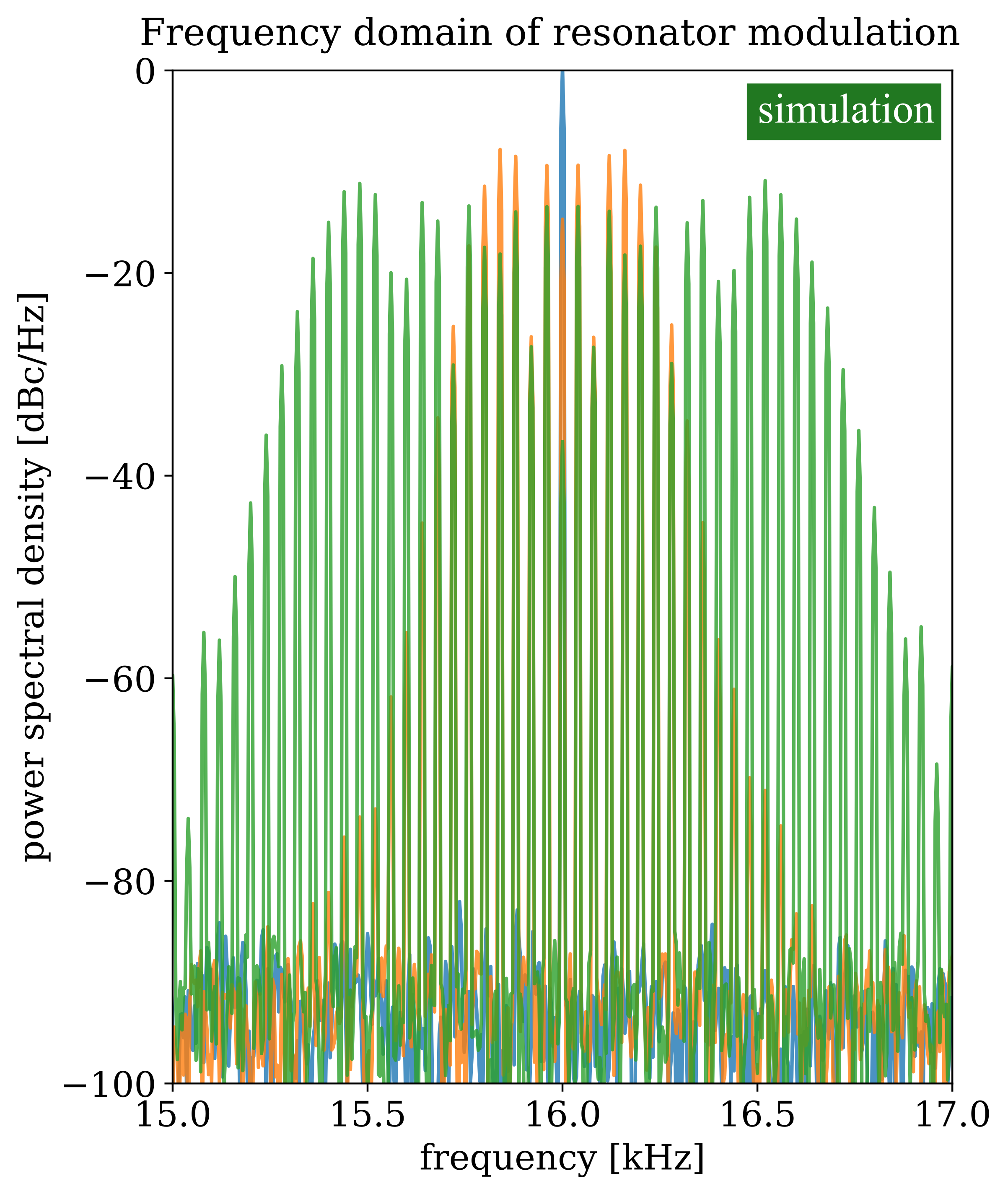}
         \label{fig:resmod_ft}
     \end{subfigure}
        \caption{(Color online) \emph{Left} Simulated time-domain representations of arbitrary input detector signals with constant frequency but varying amplitude. 
\emph{Center} Simulated time-domain representations of the flux ramp modulated resonance frequency for the same input detector signals with $4\Phi_0$, 4kHz sawtooth flux ramp. 
	\emph{Right} Frequency-domain representations of the same (simulated) resonance frequency modulation as the center panel. Note that the detector signals appear as sidebands of the $\Phi_0$ rate of the flux ramp, ie at $f_\mathrm{fr}\pm Nf_\mathrm{sig}$ for some $N$ relating to the amplitude of the detector signal and the asymmetry of the SQUID. Noise is applied to achieve about 100dB of dynamic range. }
\label{fig:umuxsig}
\end{figure}

A consequence of this flux ramp modulation is that the detector signal must be approximately constant over the course of one flux ramp cycle. 
Conversely, since the phase is typically estimated on order once per flux ramp cycle, the flux ramp rate sets the sampling rate of the detector signal. 
Fig.~\ref{fig:umuxsig} shows the time domain of a detector signal (left) and the time (center)/frequency (right) domain representation of the resonance frequency after it has been modulated by the flux ramp and detector signal. 
In particular, the right panel is the spectrum of the resonance frequency itself, not the spectrum of the transmitted rf power. 
We see that the detector signal appears as sidebands of the $\Phi_0$ rate, defined as the period of the sawtooth waveform multiplied by the number of $\Phi_0$ swept per sawtooth (typically 3-6).

\subsection{Readout of \umux}
Traditionally, microwave resonator-based detector and/or multiplexing systems are read out with warm electronics systems that send in stationary probe tones tuned to each resonance.\cite{blast16, fruitwala20}
As the resonance moves, the probe tone's amplitude and phase response are modulated, allowing for the resonance position to be inferred. 
The peak to peak frequency shift of the resonance as a result of the SQUID modulation is typically chosen to match the resonator bandwidth. 
Smaller shifts give up SQUID gain and thus degrade signal to noise, while overcoupled systems risk resonator bistability. 
Furthermore, in the case of a fixed tone readout system the resonance may shift off the probe tone entirely for some fraction of the flux ramp period, resulting in a distorted SQUID curve and/or loss of information about the resonance position.\cite{mates_thesis}

Recent development of ``tone-tracking'' algorithms have allowed for relaxation of some of these restrictions. 
The SLAC Microresonator RF (SMuRF) electronics system coupled with \umux devices typically fabricated by NIST has been demonstrated on fielded CMB systems and is being used for future large-scale observatories.\cite{ari20, so18, smurf18, dober21, ba20}  
Other projects are considering the use of SMuRF or new tone-tracking readout systems for both bolometric and calorimetric \umux applications.\cite{lynx_concept, ali21}
In the SMuRF scheme, an active feedback loop drives the probe tone to the center of the resonance, or the linear portion of the resonator phase slope. 
In order to enable flux ramping at higher rates, the SMuRF electronics rely on an adaptive filter to fit for the phase coefficients of a known sinusoidal frequency modulation.\cite{smurf18}

In these proceedings, we explore bandwidth limitations inherent to the multiplexer itself and how such limitations are impacted by the readout electronics. 

\section{\umux Bandwidth}
Since the sampling rate of a \umux system is set by the flux ramp rate, we first consider how quickly the resonance frequency itself can be tuned and read out. 
When the resonator is tuned to a new frequency, the photons at the new frequency must have time to fill and escape the resonator for the S-parameters to update.
Thus, changes in frequency as monitored through the transmission or reflection coefficients are limited by the ring-up time of the resonator $Q/\omega_0$, where $\omega_0$ is the resonance frequency.
For resonators optimized for CMB bolometric applications, we have typical $Q$s of $\sim 5\times 10^4$ and a typical resonance frequency of $f_0\sim 5$ GHz yielding a ring-up time of $\sim 10\mu\mathrm{s}$.\cite{dober21}

\begin{figure}[htbp]
\begin{center}
\includegraphics[width=0.8\textwidth]{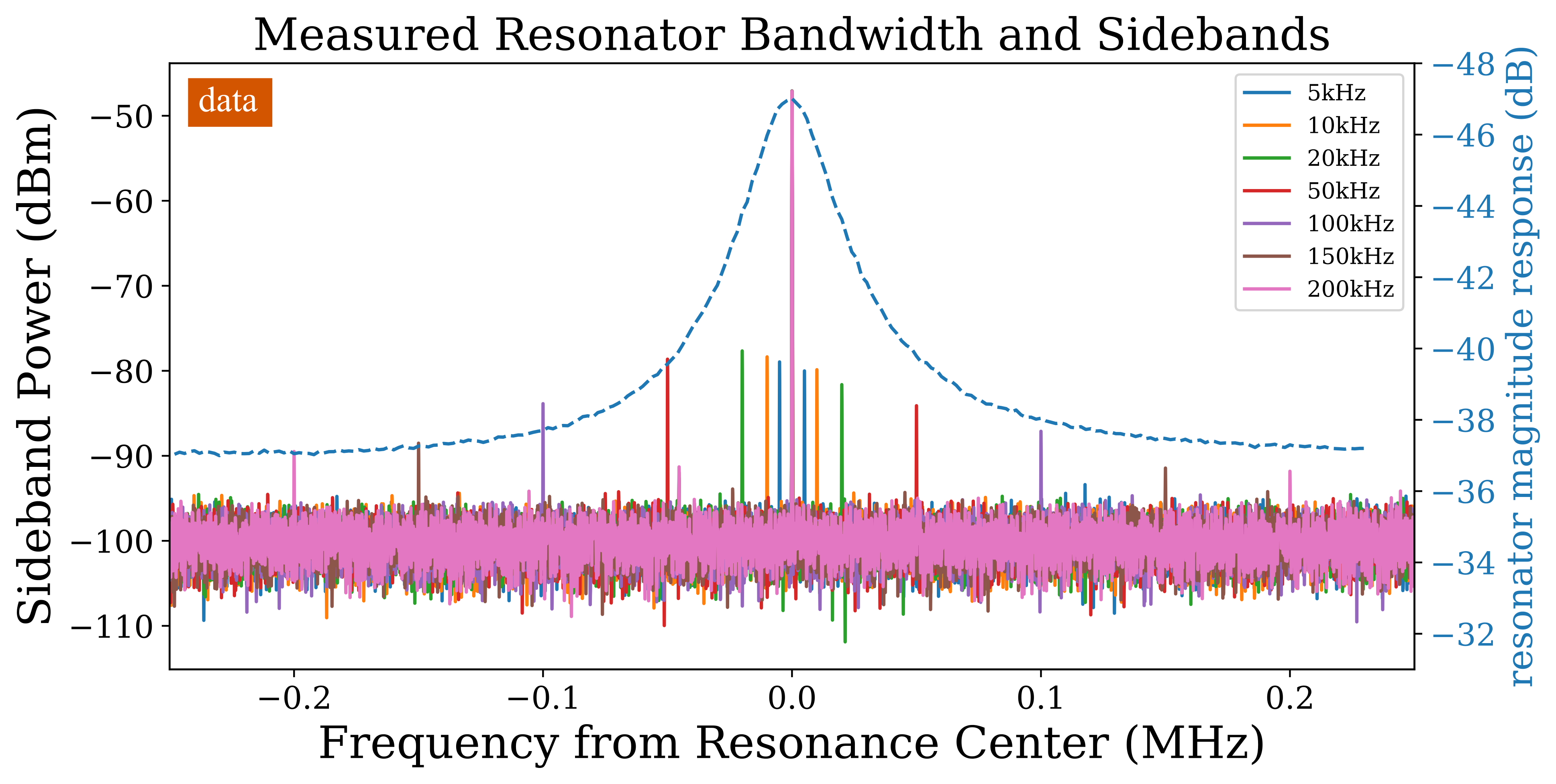}
	\caption{(Color online) Measured sideband power (left axis) of fixed amplitude signals injected on the flux ramp line of a CMB-style \umux resonator. The dashed line and right axis (blue) represent the resonance S21. The bandwidth of this resonance as defined by the FWHM is 80kHz. The resonator is probed via a fixed amplitude, fixed frequency tone tuned to the resonance frequency; thus, sidebands are the result of the probe tone photons being modulated by the changing resonance frequency. The sideband power is limited for small offset frequencies due to the finite amplitude of the injected signal, but rolls off at higher offset frequencies due to the resonance lineshape.}
\label{fig:sidebands}
\end{center}

\end{figure}

However, the resonator itself can be tuned much faster than this speed.\cite{sandberg08} 
In the case of adiabatic tuning of the resonance frequency, the photons in the resonator shift to the new resonance frequency with each cavity oscillation.  
This change can be monitored even with a fixed tone readout scheme since photons at the new resonance frequency will continue to leak out of the resonance and present as sidebands of the main probe tone. 
However, this sacrifices signal to noise because the resonator acts as a low-pass filter whose response rolls off for signal frequencies above the resonator bandwidth, as in Fig.~\ref{fig:sidebands}.\cite{zmuidzinas12}

Tone-tracking allows for modulation of the resonance to be tracked by the probe tone. 
In particular, the SMuRF adaptive filter relies on user input of the flux ramp rate to extend the bandwidth of the feedback loop to frequencies beyond the resonance bandwidth. 
Within the resonance bandwidth, this allows for improvements in linearity since the conversion between frequency shift and resonator phase remains constant. 
Near and beyond the resonator bandwidth, tracking allows for improved signal to noise due to the probe tone remaining on the responsive slope of the resonance phase slip. 
Further improvements in signal bandwidth due to the SMuRF adaptive filter relative to a fixed tone readout scheme are expected, with measurement underway.

In practice, flux ramping at higher speeds presents practical challenges beyond the theoretical limit of the resonator linewidth. 
When tone tracking, the feedback loop is ultimately limited by the warm electronics, which must fit the channelization and feedback computations within the FPGA resources. 
The adaptive feedback loop is not in principal limited by system latency, but other system clocks still practically limit the flux ramp rate achievable by the SMuRF electronics. 
This limit is exacerbated by system delays due to nonidealities such as rf cables and other wiring. 
These engineering challenges are not presently limiting the achieved flux ramp frequency but may be worth exploring for future applications where the detector count is limited by the cryogenic multiplexer. 
This is an area of active development for the SMuRF system.

\section{Additional Flux Ramp Rate Considerations}
Given the flux ramp rate sets the detector sampling rate, we now consider several factors that influence the choice of flux ramp rate when operating \umux systems. 

\subsection{Electronics Phase Noise Suppression}
Superconducting microresonators have empirically been found to exhibit excess frequency noise in phase attributed to two-level systems (TLS) in the resonator dielectric.
TLS noise often exhibits a $1/f^{1/2}$ spectrum and is observed to decrease with readout power and temperature.\cite{gao10}
In order to mitigate TLS contributions and other $1/f$-like noise in the resonator, we na\"ively prefer the fastest possible flux ramp rates that the wiring and warm electronics can support. 
However, we find that as we approach the resonator linewidth, the noise degrades due to the phase noise profile of the resonator probe tone for the existing SMuRF implementation. 

\begin{figure}[htbp]
     \centering
     \begin{subfigure}[b]{0.31\textwidth}
         \centering
         \includegraphics[width=\textwidth]{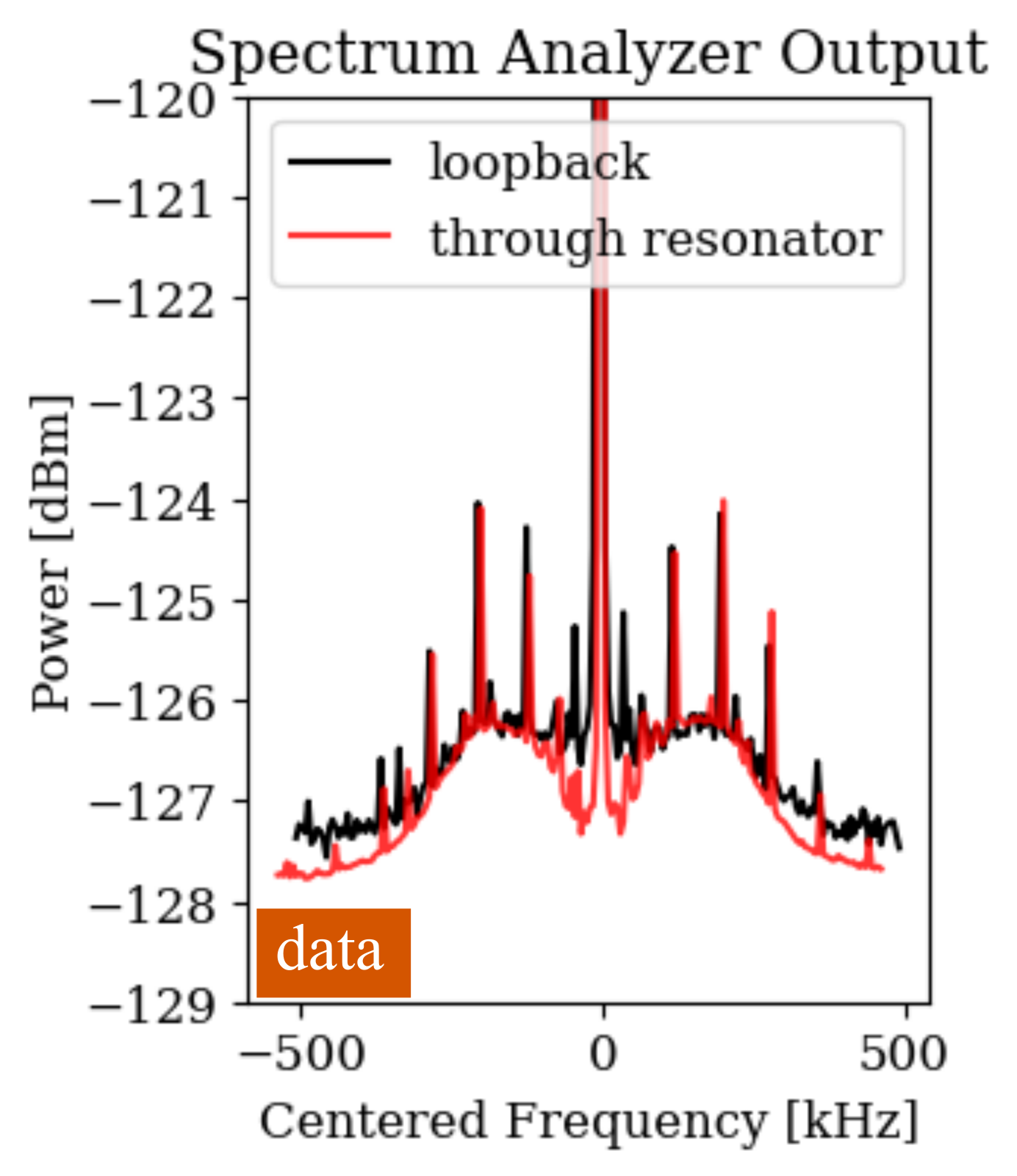}
         \label{fig:sa_wings}
     \end{subfigure}
     \hfill
     \begin{subfigure}[b]{0.36\textwidth}
         \centering
         \includegraphics[width=\textwidth]{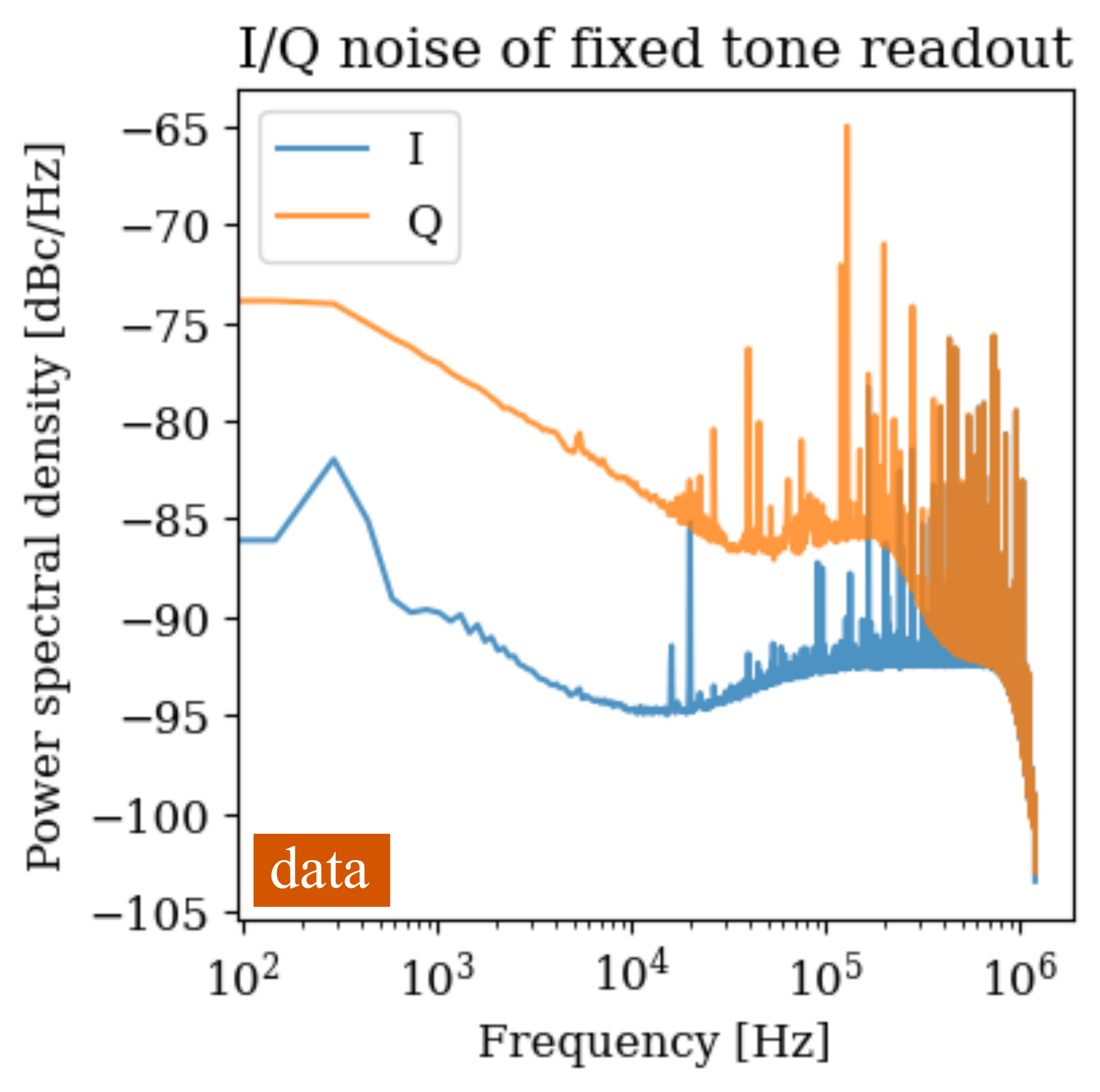}
         \label{fig:iq_psd}
     \end{subfigure}
     \hfill
     \begin{subfigure}[b]{0.28\textwidth}
         \centering
         \includegraphics[width=\textwidth]{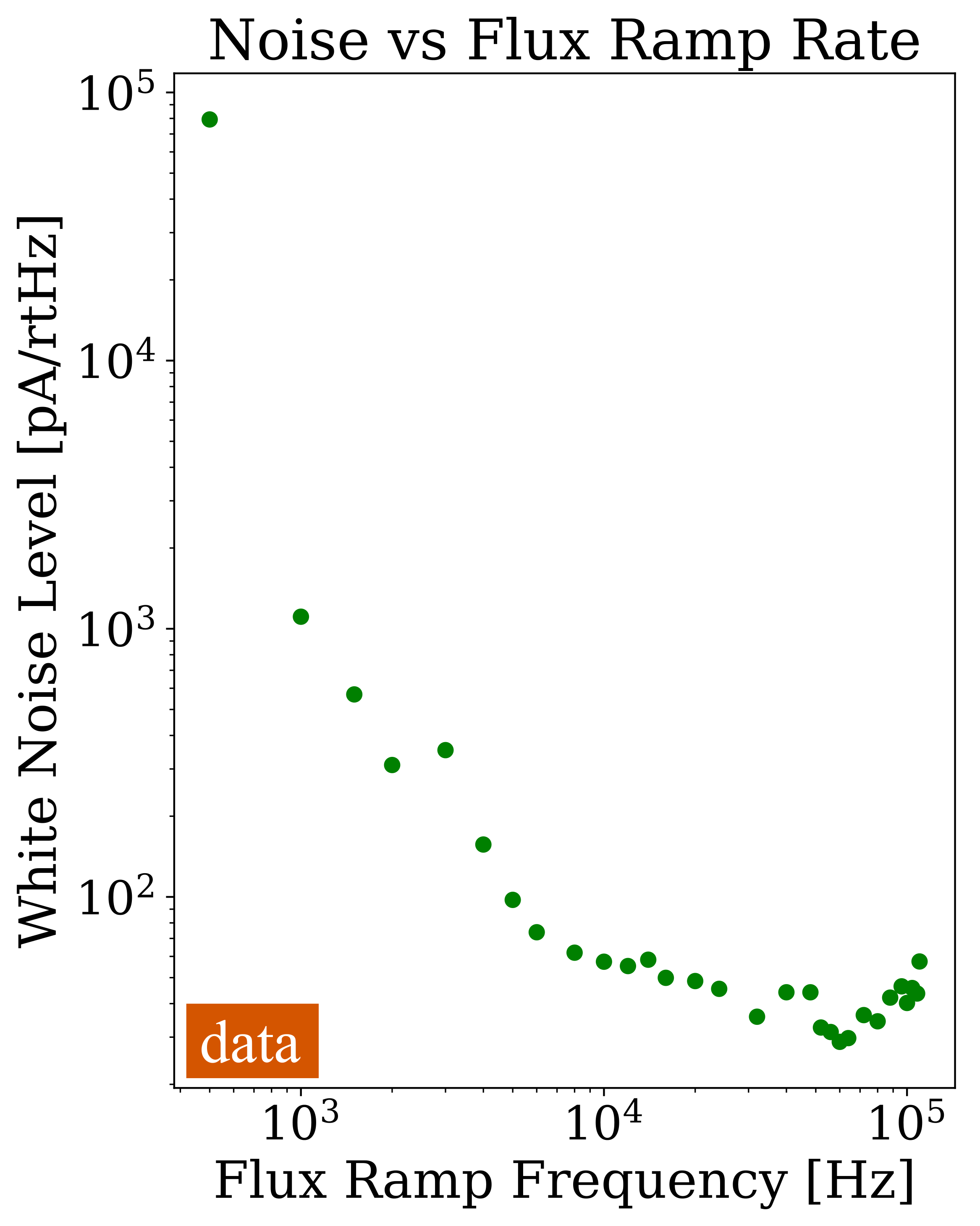}
         \label{fig:noisevfluxramp}
     \end{subfigure}
        \caption{(Color online) \emph{Left} Measured signal analyzer trace of a tone looped back without passing through the cryostat (black) versus filtered by a CMB-style resonance with bandwidth of $\sim$~80kHz (red). The local oscillator introduces a broad noise shoulder beyond the resonance bandwidth. Spurs are common mode between the two and due to the warm electronics. 
\emph{Center} Measured amplitude $(I)$ and phase $(Q)$ quadrature noise power spectral densities for the fixed tone through the resonator (red) on the left. 
\emph{Right} Measured white noise level for tone-tracked readout of the same resonance. The demodulated timestream noise profile follows a similar frequency dependence to the fixed tone phase quadrature noise profile.}
        \label{fig:noisebump}
\end{figure}

In typical resonator readout schemes, a comb of probe tones is generated by some DAC and sent down the RF feedline. 
The probe tones interact with the resonators and are read back by some ADC. 
The approach allows for cancellation of most of the probe tone phase noise. 
The resonance acts as a notch filter, attenuating noise inside its bandwidth. 
For electronics noise profiles that have some width greater than the linewidth of the resonator, however, the noise is filtered in both amplitude and phase by the resonator within its bandwidth and thus does not fully cancel on the downconversion. 
High data rate timestreams acquired with a fixed frequency probe tone on resonance thus exhibit excess noise in the spectrum corresponding to the resonance lineshape, peaking slightly above the resonance bandwidth as in the center panel of Fig.~\ref{fig:noisebump}. 
Similarly, for data with tone-tracking we see that the noise versus flux ramp rate exhibits a peak corresponding to the resonance linewidth as in the right panel of Fig.~\ref{fig:noisebump}. 

The sharpness of the bump is related to the resonance quality factor on the low frequency end and on the quality of the probe tone synthesis on the high frequency end. 
Given implementation challenges in flux ramping quickly with or without tone tracking, this practically sets a maximum flux ramp rate within the resonance bandwidth for optimal noise performance. 
Conversely, higher flux ramp rates with low noise penalty may be achieved by improving the phase noise profile of the probe tone generation. 
This is an area of expected improvement for future revisions of the SMuRF electronics.\cite{yu21}

\subsection{Aliased Noise}
\begin{figure}[htbp]
\begin{center}
     \centering
     \begin{subfigure}[b]{0.6\textwidth}
         \centering
         \includegraphics[width=\textwidth]{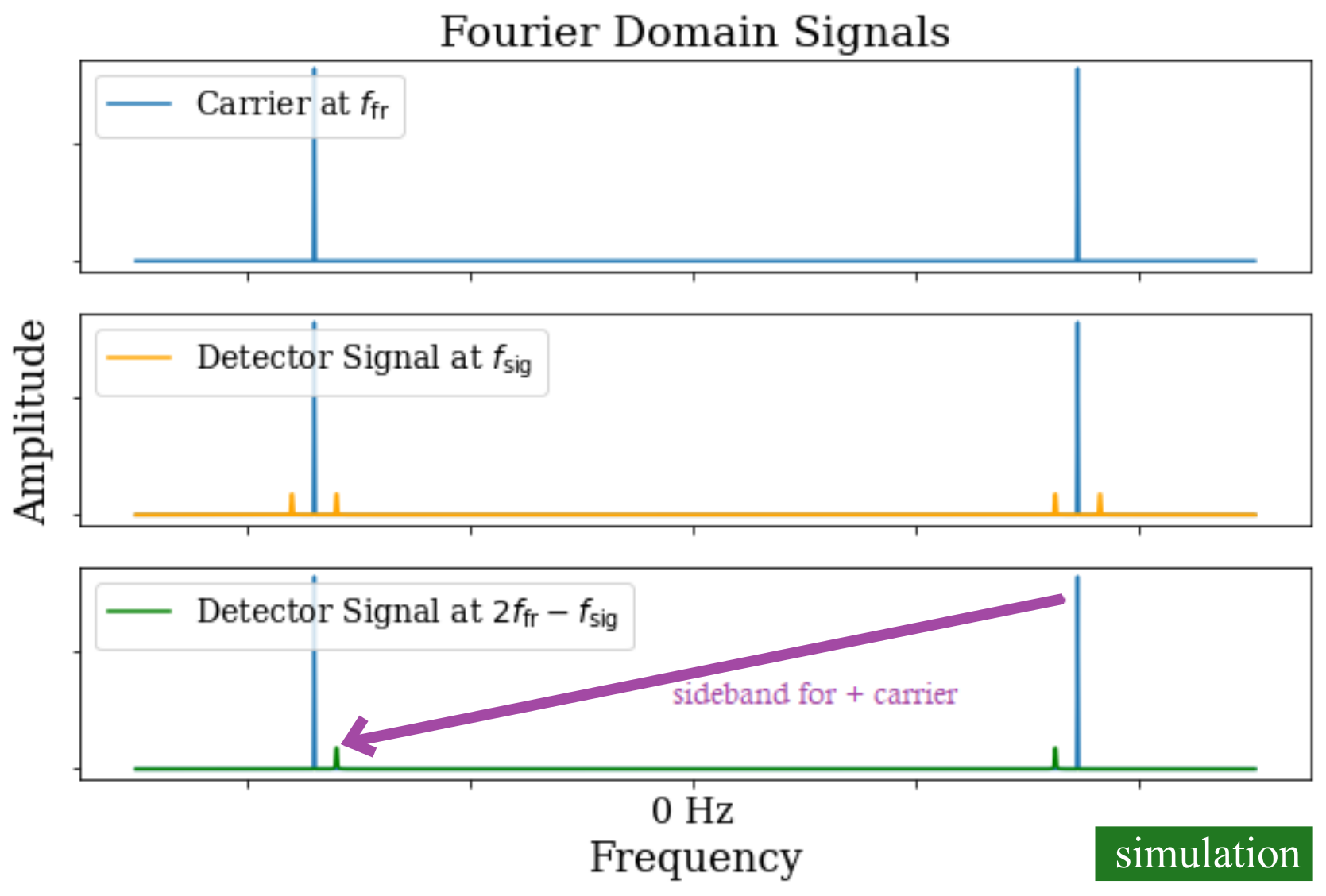}
         \label{fig:aliasing_freq}
     \end{subfigure}
     \begin{subfigure}[b]{0.33\textwidth}
         \centering
         \includegraphics[width=\textwidth]{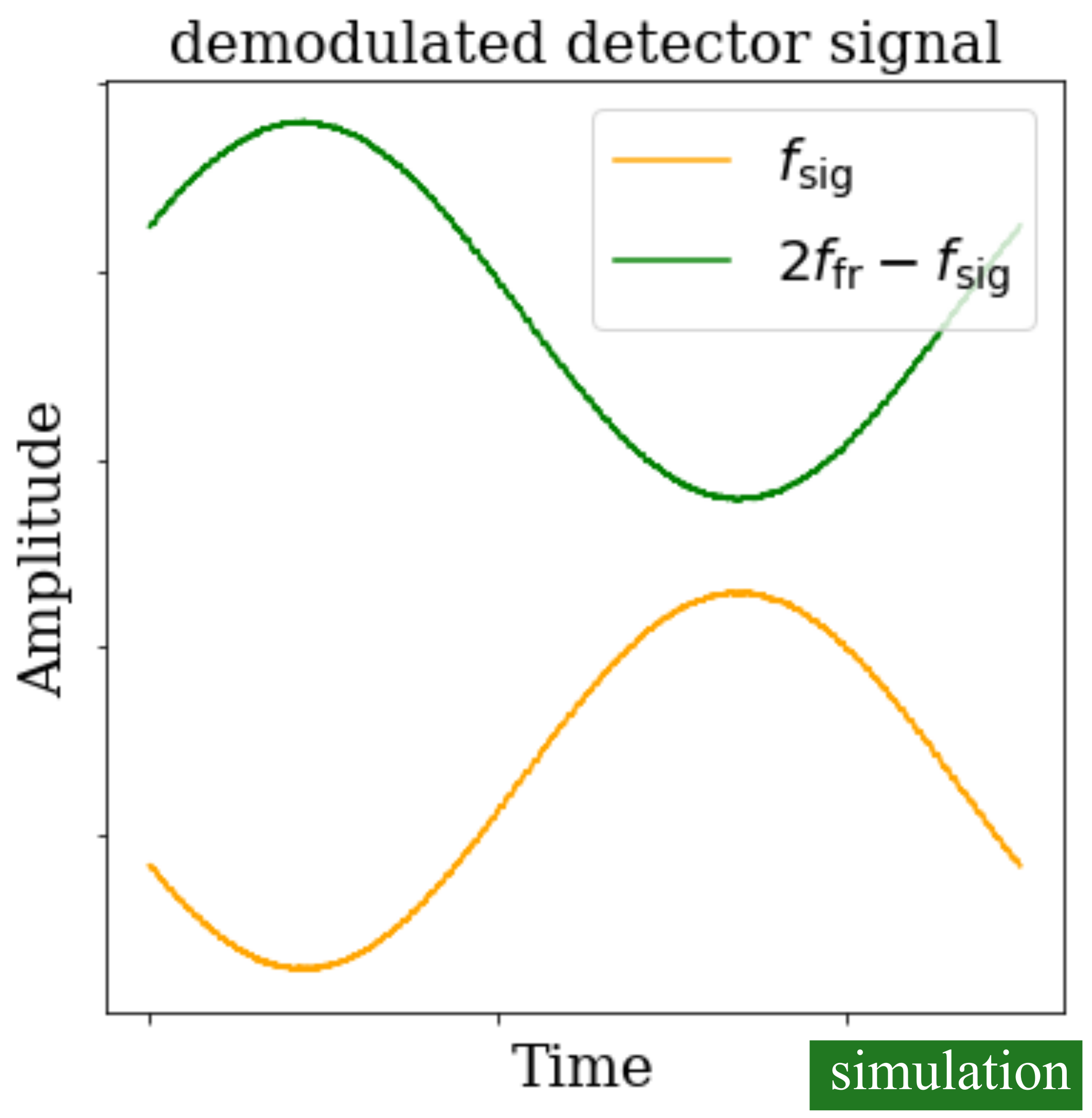}
         \label{fig:aliasing_time}
         \vspace{0.5cm}
     \end{subfigure}
        \caption{(Color online) \emph{Left} (Top) Simulated Fourier domain representation of a resonance frequency for a resonator modulated by a purely sinusoidal SQUID. The modulation appears as positive and negative peaks at the carrier rate $f_\mathrm{fr}$. (Middle) The same simulated flux ramp modulated frequency with an additional sinusoidal detector signal at $f_\mathrm{sig}$ injected. The detector signal appears as sidebands of the flux ramp carrier frequency. (Bottom) The same simulated flux ramp modulated frequency with an additional sinusoidal detector signal at $2f_\mathrm{fr}-f_\mathrm{sig}$ injected. The arrow denotes the negative signal sideband of the positive flux ramp carrier, which looks like the positive signal sideband of the negative flux ramp carrier. 
        \emph{Right} Simulated flux ramp demodulated detector signals in the time domain for the same signal frequencies as in the left panel. Up to a phase offset, they are indistingushable from each other.}
\label{fig:umuxsig_alias}
\end{center}
\end{figure}

A separate consideration for flux ramp rate is the aliasing of flux ramp harmonic sidebands into the signal band. 
Due to the phase modulation scheme of \umux, the detector signal at some frequency $f_\mathrm{sig}$ is modulated by the SQUID's quasi-sinusoidal response at some flux ramp rate $f_\mathrm{fr}$. 
In the cleanest possible case with a purely sinusoidal $V-\Phi$ relation the signal thus appears as sidebands at  $f_\mathrm{fr}\pm f_\mathrm{sig}$; non-sinusoidal or asymmetric SQUID curves have more nontrivial sideband structure at multiples of $f_\mathrm{sig}$. 
These sidebands are indistinguishable from sidebands of $Nf_\mathrm{fr} - f_\mathrm{sig}$ for arbitrary $N$. 
In Fig.~\ref{fig:umuxsig_alias} we show the Fourier domain representation of a flux ramp carrier $f_\mathrm{fr}$ modulated by detector signals at $f_\mathrm{sig}$ versus $2f_\mathrm{fr}-f_\mathrm{sig}$ (left) and the time domain representation of the flux ramp demodulated detector signals (right) illustrating this effect. 
This aliasing occurs independent of the readout electronics scheme or asymmetry of the SQUID and is inherent to the phase modulation scheme of \umux. 

Thus if the readout electronics are capable of supporting it, faster flux ramp rates avoid aliasing in excess noise at harmonics of the flux ramp frequency due to noise at higher harmonics being rolled off by the resonator. 
For typical CMB bolometers, the thermal and electrical time constants both roll off the photon noise from the sky  quite aggressively and this constraint is not especially strong, but future systems must consider this limit when setting the bandwidth of detectors interfacing with \umux readout. 
In the case of calorimetry where the pulsed signal contains spectral information out to higher frequencies, this poses a problem for any schemes that undersample the full rising edge of the pulse.

\section{Conclusion}
Here we consider several theoretical and practical limitations on bandwidth of the microwave SQUID multiplexer, a cryogenic multiplexing scheme that has been used for both bolometric and calorimetric applications. 
The common flux ramp, which linearizes the SQUID response and eliminates the need for per-detector feedback lines, sets the bandwidth of the system. 
While the resonator itself can be tuned arbitrarily quickly, the linearity and signal to noise ratio degrade with the resonance lineshape unless a mitigation scheme such as tone-tracking electronics are employed. 
Two-level systems noise and aliasing further push the desired flux ramp rate to higher frequencies. 
However, practical considerations including warm electronics probe tones noise profiles, cable delays, and FPGA resource limits often constrain the achievable flux ramp rate; several development paths for improving the bandwidth through readout electronics are discussed. 
These together suggest that future applications may optimize tone-tracking electronics to achieve comparable signal-to-noise with narrower resonances, thus paving the way for higher multiplexing factors and higher detector counts.

\begin{acknowledgements}
The authors thank Kent Irwin for useful discussions. 
CY was supported in part by the National Science Foundation Graduate Research Fellowship Program under Grant No. 1656518.
\end{acknowledgements}

\noindent \small \textbf{Data Availability} 
The datasets used to generate the plots and conclusions in these proceedings are available from the corresponding author on reasonable request.

\pagebreak

\end{document}